\documentclass[preprint,10pt,5p]{elsarticle}

\usepackage{amssymb}
\usepackage{mathrsfs}

\usepackage{hyperref}

\usepackage{graphicx}
\usepackage{dcolumn}
\usepackage{bm}
\usepackage{mathrsfs}
\usepackage{multirow}
 
\usepackage{atlasphysics}

\usepackage{lmetdefs}

\begin{document}

\journal{Physics Letters B}

\begin{frontmatter}

\title{
\vspace{-0.8cm} \flushright{\normalsize{CERN-PH-EP-2011-121}} \\ \center{
  Search for a heavy gauge boson decaying to a charged lepton and a neutrino
  in 1~\ifb\ of $pp$ collisions at $\sqrt{s}= 7 \tev$ using the ATLAS detector}
}

\author{The ATLAS Collaboration}

\begin{abstract}
The ATLAS detector at the LHC is used to search for heavy charged gauge bosons (\wp),
decaying to a charged lepton (electron or muon) and a neutrino. 
Results are presented based on the analysis of \pp\ collisions at a center-of-mass energy of
7~\tev\ corresponding to an integrated luminosity of 1.04~\ifb.
No excess above Standard Model expectations is observed.
A \wp\ with Sequential Standard Model couplings is excluded at the 95\% confidence level for masses up to 2.15~\tev. 
\end{abstract}

\end{frontmatter}

\section{Introduction}

The high-energy collisions at the CERN Large Hadron Collider provide new opportunities
to search for physics beyond the Standard Model (SM) of strong and electroweak interactions.
One extension common to many models is the existence of additional heavy gauge
bosons~\cite{pdg:2010}, the charged ones commonly denoted \wp.
Such particles are most easily searched for in their decay to a charged lepton (electron or muon)
and a neutrino.

This letter describes such a search performed using
7~\tev\ \pp\ collision data collected with the ATLAS detector during 2011 and corresponding to a total
integrated luminosity of 1.04~\ifb.
No \wp\ signal is observed, and the data
are used to extend current limits~\cite{cdf:Wprime2010, cms:wpmu2011mar, atlas:wprime_2010_pub}
on \xbr\ (cross section times branching fraction) as a function of \wp\ mass.
The significant improvement over the previous ATLAS result~\cite{atlas:wprime_2010_pub} comes
mostly from the increase in available integrated luminosity, but also reflects optimization of
the event selection and increased acceptance in the muon channel.
A lower limit on the mass of a \wp\ boson in the Sequential Standard Model (SSM),
i.e.\ the extended gauge model of Ref.~\cite{ssm} with \wp\ coupling to $WZ$ set to zero, is also reported.
In this model, the \wp\ has the same couplings to fermions as the SM \w~boson and
thus a width which increases linearly with the \wp\ mass.

The analysis presented here identifies candidates in the electron and muon channels
and sets separate limits for \mbox{\wpe} and \wpmu.
In addition, combined limits are evaluated, assuming the same branching fraction for both channels.
The kinematic variable used to identify the \wp\ is the transverse mass
\begin{equation}
\mt = \sqrt{ 2 \pt \met (1 - \cos \varphi_{l\nu})},
\end{equation}
which displays a Jacobian peak that falls sharply above the resonance mass.
Here \pt\ is the lepton transverse momentum, \met\ is the magnitude of the missing transverse momentum (\mettext),
and $\varphi_{l\nu}$ is the angle between the \pt\ and \mettext\ vectors.
Throughout this letter, transverse refers to the plane perpendicular to the colliding beams, longitudinal
means parallel to the beams, \mytheta\ and \myphi\ are the polar and azimuthal angles with respect to the
longitudinal direction, and pseudorapidity is defined as $\eta = -\ln(\tan(\theta/2))$.

The main background to the \wpl\ signal comes from the high-\mt\ tail of SM \w~boson decay to the same final state.
Other backgrounds are \z\ bosons decaying into two leptons where one lepton is not reconstructed,
\w\ or \z\ decaying to \tauleps\ where a $\tau$ subsequently decays to an electron or muon, and
diboson production. These are collectively referred to as the electroweak (EW) background.
In addition, there is a background contribution from \ttbar\ production which is most important for the lowest
\wp\ masses considered here, where it constitutes about 10\% of the background after event selection.
Other strong-interaction background sources, where a light or heavy hadron decays semileptonically or a jet
is misidentified as an electron, are estimated to be at most 10\% of the total background in the electron channel
and a negligible fraction in the muon channel, again after final selection.
These are called QCD background in the following.

\section{Data}

The ATLAS detector~\cite{atlas:detector} has three major components: the inner tracking detector, 
the calorimeter and the muon spectrometer.
Charged particle tracks and vertices are reconstructed with silicon pixel and silicon strip detectors
covering $|\eta|<2.5$ and transition radiation detectors covering $|\eta| < 2.0$,
all immersed in a homogeneous 2~T magnetic field provided by a superconducting solenoid.
This tracking detector is
surrounded by a finely-segmented, hermetic calorimeter system that covers \mbox{$|\eta| < 4.9$}
and provides three-dimensional reconstruction of particle showers.
It uses liquid argon for the inner, electromagnetic
compartment followed by a hadronic compartment based on scintillating tiles in the central region ($|\eta| < 1.7$)
and additional liquid argon for higher $|\eta|$.
Outside the calorimeter, there is a muon spectrometer with air-core toroids providing a magnetic
field, whose integral averages about 3~Tm.
The deflection of the muons in the magnetic field is measured with three layers of precision drift-tube chambers for
$|\eta| < 2.0$ and one layer of cathode-strip chambers followed by two layers of drift-tube chambers
for $2.0 < |\eta| < 2.7$.
Additional resistive-plate and thin-gap chambers provide muon triggering capability and measurement of
the $\varphi$ coordinate.

\begin{sloppypar}
The data used in the electron channel are the events recorded with a trigger requiring the presence of an
electron with $\pt > 20\gev$.
The efficiency of this trigger is 98\%.
For the muon channel, matching tracks in the muon spectrometer and inner detector with combined $\pt > 22\gev$ are
used to identify events. Events are also recorded if a muon with $\pt>40\gev$ is found in the muon spectrometer.
The muon trigger efficiency is 80-90\% in the regions of interest.
\end{sloppypar}

Each energy cluster reconstructed in the electromagnetic compartment of the calorimeter with \mbox{$\et > 25\gev$}
and $|\eta| < 1.37$ or $1.52 < |\eta| < 2.47$
is considered as an electron candidate if it matches with an inner detector track.
The electron direction is defined as that of the reconstructed track and its energy as that of the cluster,
with a small (less than 2\%) \eta-dependent energy scale correction.
The resolution of the energy measurement is 2\% for $\et \approx 50\gev$ and approaches 1\% in the high-\et\
range relevant to this analysis.
To discriminate against hadronic jets,
requirements are imposed on the lateral shower shapes in the first two layers
of the electromagnetic part of the calorimeter
and the fraction of energy leaking into the hadronic compartment.
A hit in the first pixel layer is required to reduce background from photon conversions in the
inner detector material.
These requirements give about 90\% identification efficiency for electrons with \mbox{$\et > 25 \gev$}
and a $2\times10^{-4}$ probability to falsely identify jets as electrons before isolation requirements are
imposed~\cite{atlas:wz2010}.

Muon tracks can be reconstructed independently in both the inner detector and muon spectrometer, and
the muons used in this study are required to have matching tracks in both systems.
The muons are required to have $\pt > 25\gev$, where the
momentum of the muon is obtained by combining the inner detector and muon spectrometer measurements.
To ensure precise measurement of the momentum,
muons are required to have hits in all three muon layers and
are restricted to those \eta-ranges where the muon spectrometer alignment is best
understood: approximately $|\eta| < 1.0$ and $1.3 < |\eta| < 2.0$.
The average momentum resolution is currently about 15\% at $\pt = 1\tev$.
About 80\% of the muons in these \eta-ranges are reconstructed,
with most of the loss coming from regions with limited detector coverage.

The \mettext\ in the electron channel is obtained from a vector sum over calorimeter cells associated
with topological clusters and using local hadronic calibration~\cite{atlas:csc}:
\begin{equation}
\vecmet = \vecmetcalo = - \vecsumettopo.
\label{eqn:metcal}
\end{equation}
The topological clusters reduce contributions from electronic noise.
The \et\ of cells associated with the electron is corrected so their sum equals the electron \et.
Muons only deposit a small fraction of their energy in the calorimeter, and so, in the muon channel,
the \mettext\ is obtained from
\begin{equation}
\vecmet = \vecmetcalo - \vecptmu + \vecetlossmu.
\label{eqn:metcalmu}
\end{equation}
The second term in this vector sum subtracts the muon transverse momentum and the last corrects for
the transverse component of the energy deposited in the calo\-rimeter by the muon, which is included in both
of the first two terms.
The energy loss is estimated by integrating the amount of material traversed and applying a calibrated
conversion from path length to energy for each material type.

This analysis makes use of all the $\sqrt{s} = 7\tev$ data collected in March-June 2011 that satisfy data quality
requirements which guarantee the relevant detector systems were operating properly.
The integrated luminosity for the data used in this study is 1.04~\ifb\ in both the electron and muon
decay channels.
The uncertainty on this estimate is 3.7\%.

\section{Simulation}

Except for the QCD background, which is estimated from data, expected signal and background levels are evaluated with
simulated samples and normalized using calculated cross sections and the integrated luminosity of the data.

\begin{sloppypar}
The \wp\ signal and
the \wzbg\ boson backgrounds
are generated with
\pythia~6.421~\cite{pythia} using MRST LO*~\cite{mrst} parton distribution functions (PDFs).
The \ttbar\ background is generated with \mcatnlo~3.41~\cite{mcatnlo}.
For all samples, final-state photon radiation is handled by \photos~\cite{photos}.
ATLAS full detector simulation~\cite{atlas:sim} based on \geant4~\cite{geant} is used to
propagate the particles and account for the response of the detector.

The \pythia\ signal model for \wp\ has \mbox{$V-A$} SM couplings but does not include
interference between \w\ and \wp. Decays to channels other than $e\nu$ and $\mu\nu$, including $\tau\nu$,
$ud$, $sc$ and $tb$ are included in the calculation of the \wp\ widths but are not explicitly
included as signal or background.
At high mass ($\mwp > 1~\tev$),
the branching fraction to any of the lepton decay channels is 8.2\%.

The \wlnu\ events are reweighted to have the NNLO (next-to-next-to-leading-order QCD) mass dependence of ZWPROD~\cite{zwprod}
with MSTW2008 PDFs~\cite{mstw} and following the $G_{\mu}$ scheme~\cite{horace}.
Higher-order electroweak corrections (in addition to the photon radiation included in the simulation)
are calculated using \horace~\cite{horace, horaceNC}.
In the high-mass region of interest, the electroweak corrections reduce the cross sections
by 11\% at $\mlnu = 1\tev$ and by 18\% at $\mlnu = 2\tev$.
\end{sloppypar}

The \wlnu\ and \zll\ cross sections are calculated at NNLO 
using FEWZ~\cite{fewz,Gavin:2010az} with the same PDFs, scheme and electroweak corrections used in
the ZWPROD event reweighting.
The \wpl\ cross sections are calculated in the same way, except
the electroweak corrections beyond final-state radiation are not included because the calculation for the SM \w\
cannot be applied directly.
The \ttbar\ cross section is calculated at approximate-NNLO~\cite{Moch:2008qy, Langenfeld:2009tc, Aliev:2010zk}
assuming a top-quark mass of 172.5~\gev.
The signal and most important background values for \xbr\ are listed in Table~\ref{tab:xsec}.

\begin{sloppypar}
Cross-section uncertainties for \wpl\ and the \wzbg~\cite{atlas:wz2010} and \ttbar~\cite{atlas:ttbar2010}
backgrounds are estimated from the MSTW2008 PDF error sets,
the difference between MSTW2008 and CTEQ6.6~\cite{cteq} PDF sets, and variation of renormalization and factorization scales
by a factor of two.
The estimates from the three sources are combined in quadrature.
Most of the net uncertainty comes from the error sets and the MSTW-CTEQ difference, in roughly equal proportion.
The uncertainty on the cross section for the \wlnu\ background varies from 5\% at $\mlnu = 500\gev$ to 19\% at $\mlnu = 2500\gev$.
\end{sloppypar}

\begin{table}[!tb]
\caption{
Calculated values of \xbr\ for \wpl\ and the leading backgrounds.
The value for $\ttbar\rightarrow\ell X$ includes all final states with at least one
lepton ($e$, $\mu$ or $\tau$).
The others are exclusive and are used for both $\ell=e$ and $\ell=\mu$.
All calculations are NNLO except \ttbar\ which is approximate-NNLO.
}
\label{tab:xsec}
\begin{center}
\begin{tabular}{l|c|l}
\hline
\hline
        & Mass        &             \\
Process &       [\gev] & $\xbr$ [pb] \\
\hline
\multirow{10}{*}{\wpl} &
   \phantom{0}500 & \phantom{000}17.25    \\
 & \phantom{0}600 & \phantom{0000}8.27    \\
 & \phantom{0}750 & \phantom{0000}3.20    \\
 & \phantom{}1000 & \phantom{0000}0.837   \\
 & \phantom{}1250 & \phantom{0000}0.261   \\
 & \phantom{}1500 & \phantom{0000}0.0887  \\
 & \phantom{}1750 & \phantom{0000}0.0325  \\
 & \phantom{}2000 & \phantom{0000}0.0126  \\
 & \phantom{}2250 & \phantom{0000}0.00526 \\
 & \phantom{}2500 & \phantom{0000}0.00234 \\
\hline
\wlnu\ & & \phantom{}10460 \\
\zgll\               & & \multirow{2}{*}{\phantom{00}989} \\
($m_{\zg}>60 \gev$)  & & \\
 \ttbarl &  & \phantom{000}89.4 \\
\hline
\hline
\end{tabular}
\end{center}
\end{table}

\section{Event selection}

Events are required to have their primary vertex reconstructed from at least three tracks with
$\pt > 0.4$~\gev\ and longitudinal distance less than 200~mm from the center of the
collision region.
Due to the high luminosity, there were typically five additional interactions per event and
the primary vertex is defined to be the one with the highest summed track $\pt^2$.
Spurious tails in \mettext\ arising from calorimeter noise and other detector problems are suppressed by checking the quality
of each reconstructed jet and discarding events where any jet has a shape indicating such problems,
following Ref.~\cite{atlas:plhc:jet_cleaning}.
Events are required to have exactly one candidate electron or one candidate muon satisfying
the requirements described above.
In addition, the inner detector track associated with the electron or muon is required to be compatible
with originating from the primary vertex, specifically to have transverse distance of closest approach $|d_0|<1$~mm
and longitudinal distance at this point $|z_0|<5$~mm.

To suppress the QCD background, the lepton is required to be isolated. In the electron channel,
the isolation energy is measured with the calorimeter in a cone $\Delta R < 0.4$
($\Delta R \equiv \sqrt{(\Delta \eta)^2 + (\Delta \varphi)^2}$) around the electron track,
and the requirement is \mbox{$\sumet < 9 \gev$,}
where the sum includes all calorimeter energy clusters in the cone excluding the core energy deposited by the electron.
The sum is  
corrected to account for additional interactions and leakage of the electron energy outside this core.
In the muon channel, the isolation energy is measured using inner detector tracks with $\pttrack > 1$~\gev\ in a cone
$\Delta R < 0.3$ around the muon track.
The isolation requirement is
\mbox{$\sumpttrack < 0.05~\pt$},
where the muon track is excluded from the sum.
The scaling of the threshold with the muon \pt\ 
reduces efficiency losses due to radiation from the muon at high \pt.

Finally, \mettext\ requirements are imposed to further suppress the QCD background.
In both channels, a fixed threshold is applied:
\mbox{$\met > 25 \gev$}.
In the electron channel, where hadronic jets may be misidentified as electrons,
a threshold proportional to the electron \et\ is also applied:
\mbox{$\met > 0.6~\et$}.

In the electron channel, the QCD background is estimated from data using the ABCD
technique~\cite{atlas:photon2010} with the isolation energy and \mettext\ serving as discriminants.
Consistent results are obtained using the ``inverted isolation'' technique described
in Ref.~\cite{atlas:wprime_2010_pub}.
In the higher mass bins ($\mt > 700\gev$) where no events remain in the estimate, the QCD background
level is set to zero and assigned an uncertainty equal to 10\% of the total background level,
a conservative upper limit based on the QCD contribution to the electron \mt\ distribution.

The QCD background for the muon channel is evaluated using a non-isolated data sample following the same
procedure used for the 2010 analysis~\cite{atlas:wprime_2010_pub}.
With the higher statistics now available, it is clear this background is less than 1\% of the total
background, so it is neglected in the following.

The same reconstruction and event selection are applied to both data and simulated samples.
Figure~\ref{fig:final_mt} shows the \pt, missing \et, and \mt\ spectra for each channel after event selection
for the data, for the expected background, and for three examples of \wp\ signals at different masses.
The agreement between the data and expected background is good.
Table~\ref{tab:nbg} shows as an example how different sources contribute to the background
for $\mt>891 \gev$, the region used to search for a \wp\ with a mass of 1500~\gev.
The \wlnu\ background dominates.
The \zll\ background is much larger in the muon channel because most of the energy of the undetected
muon is not captured in the calorimeter.

\begin{figure*}[]
  \centering
  \includegraphics[width=0.48\textwidth]{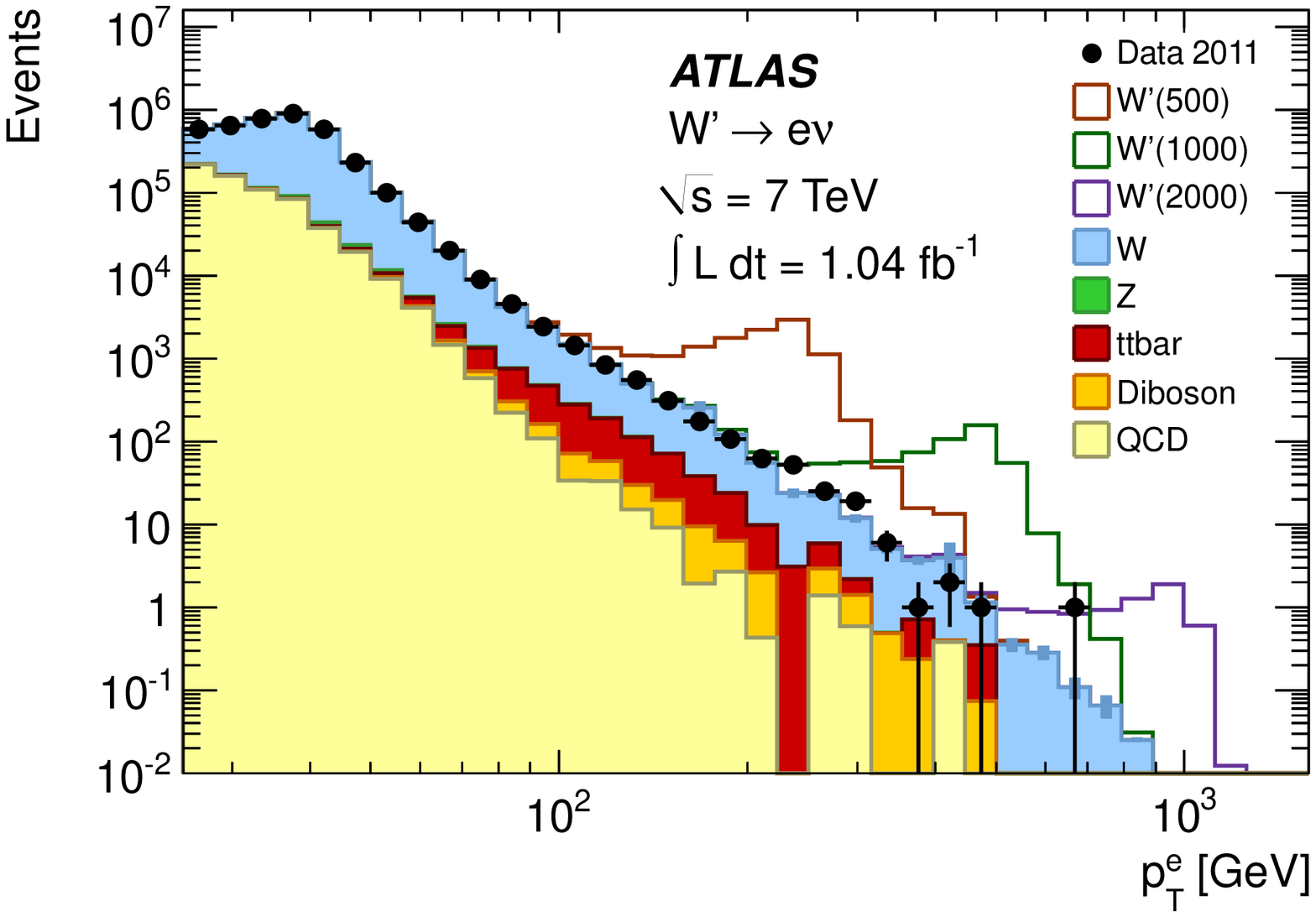}
  \includegraphics[width=0.48\textwidth]{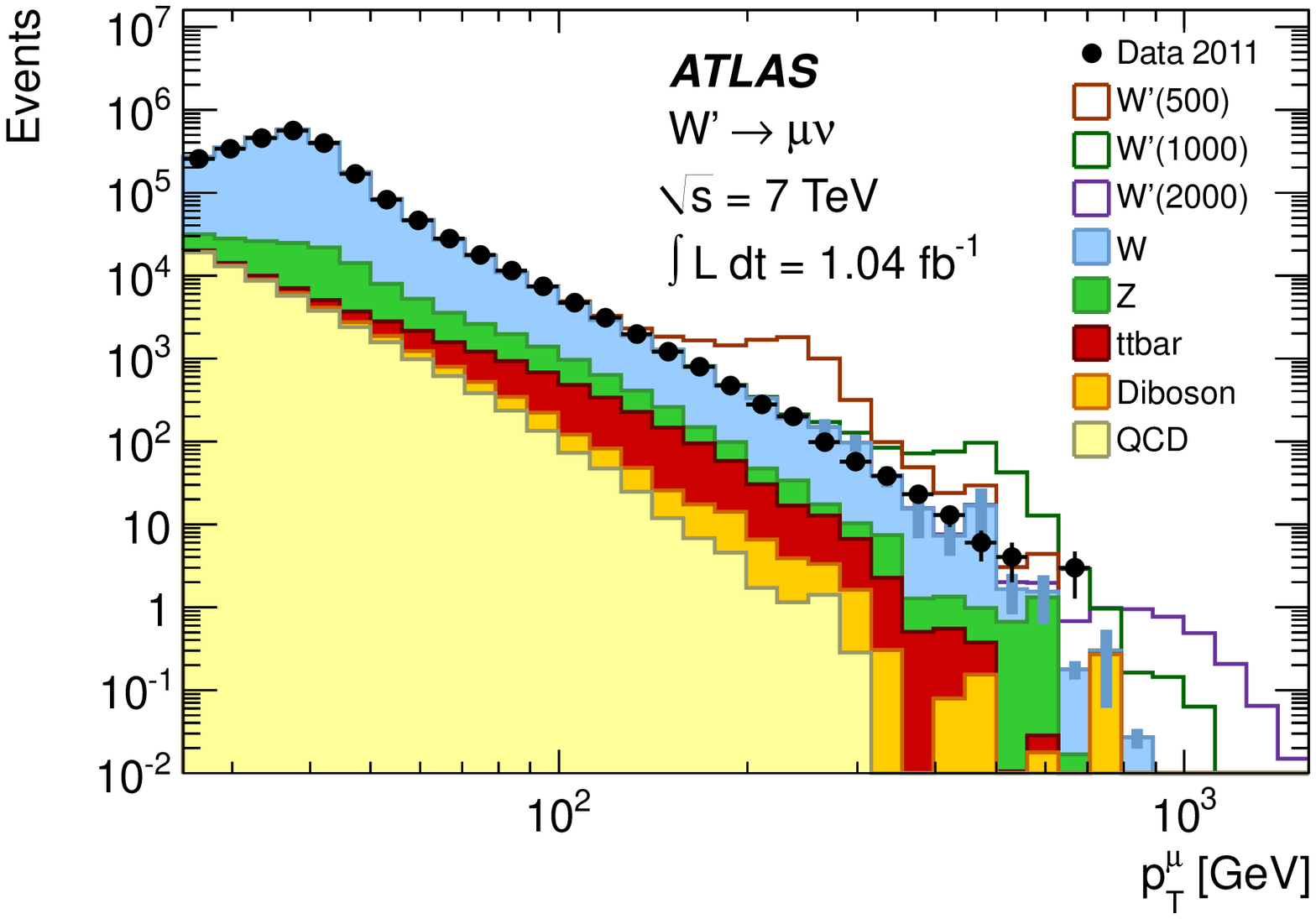}
  \includegraphics[width=0.48\textwidth]{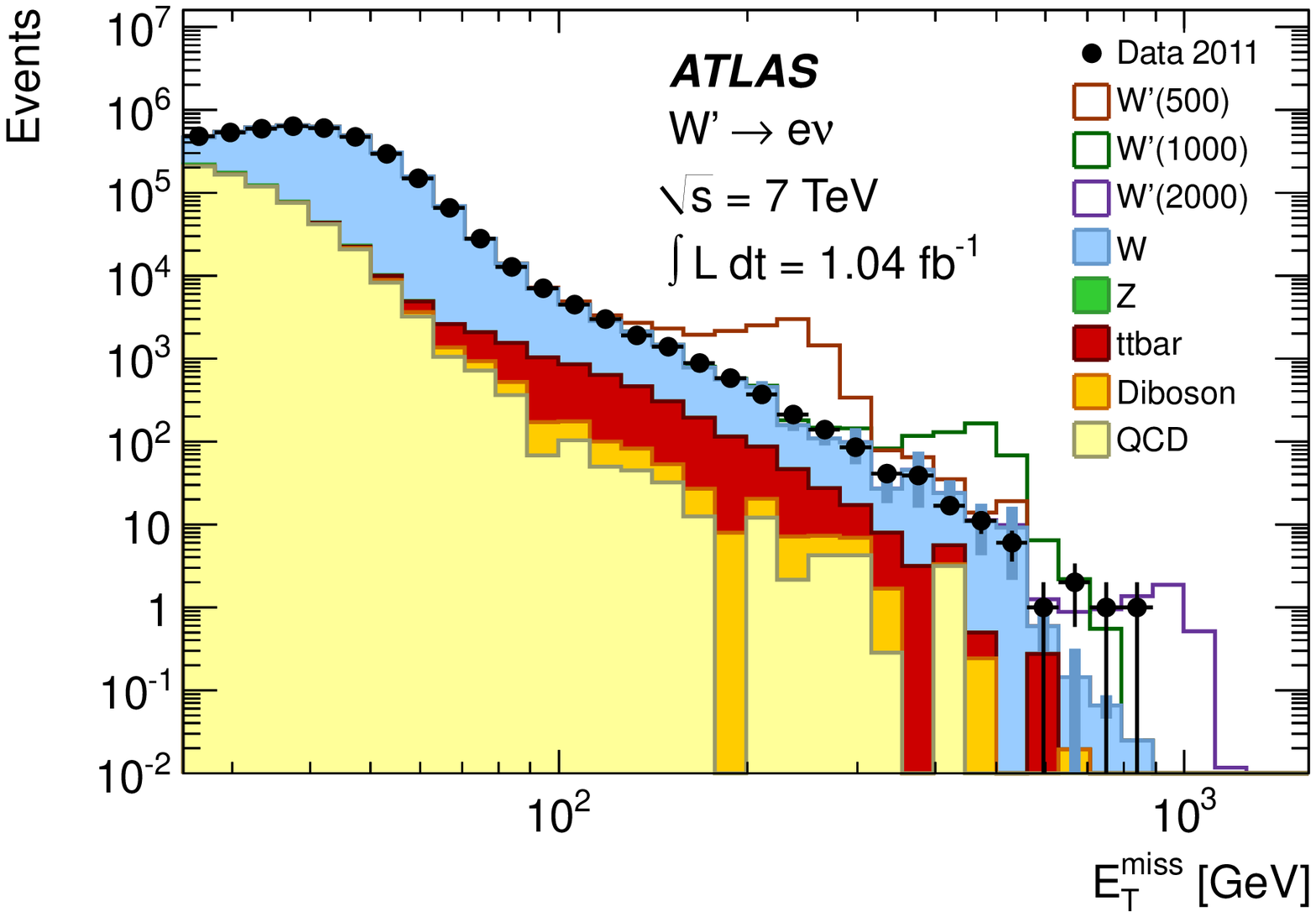}
  \includegraphics[width=0.48\textwidth]{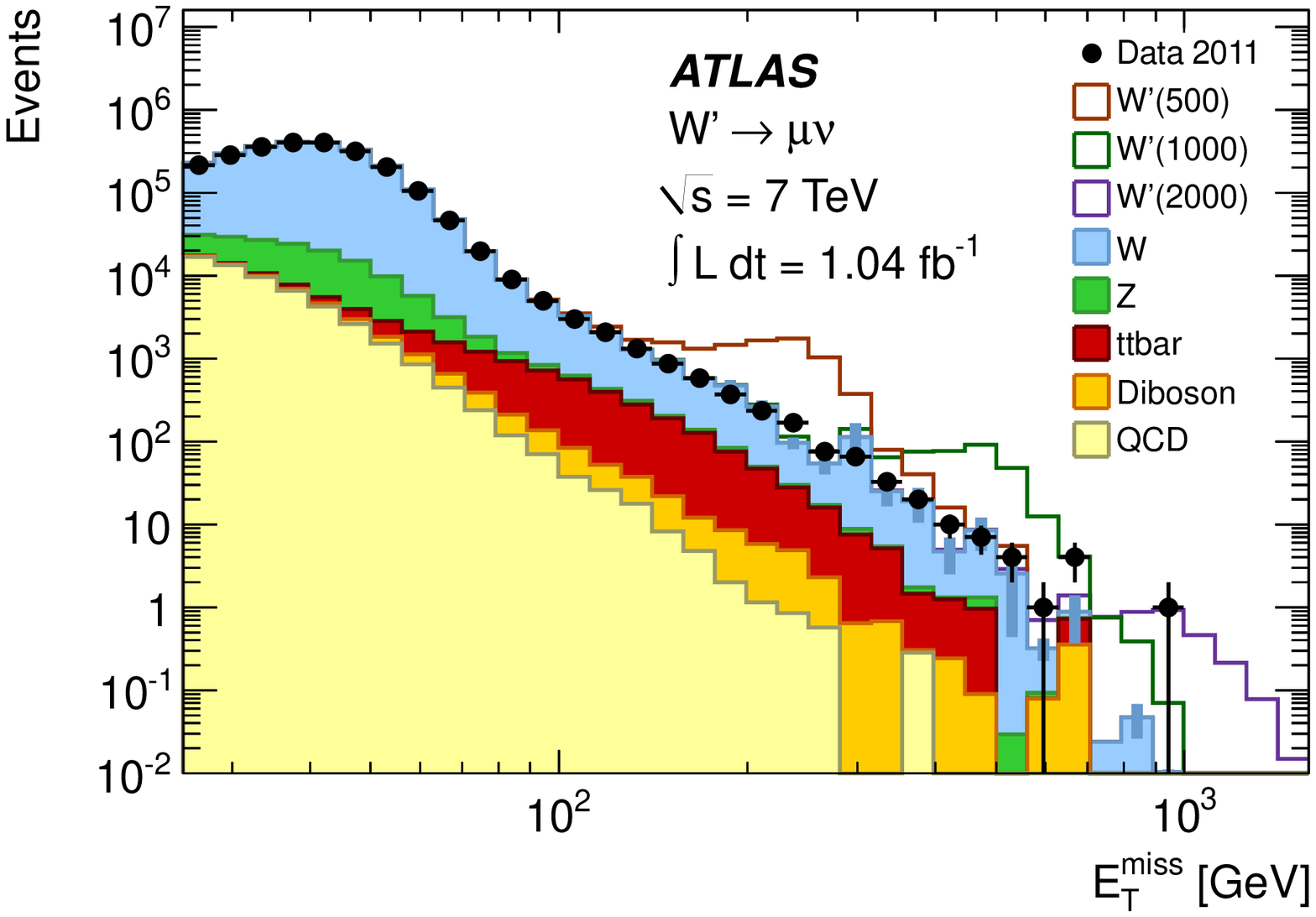}
  \includegraphics[width=0.48\textwidth]{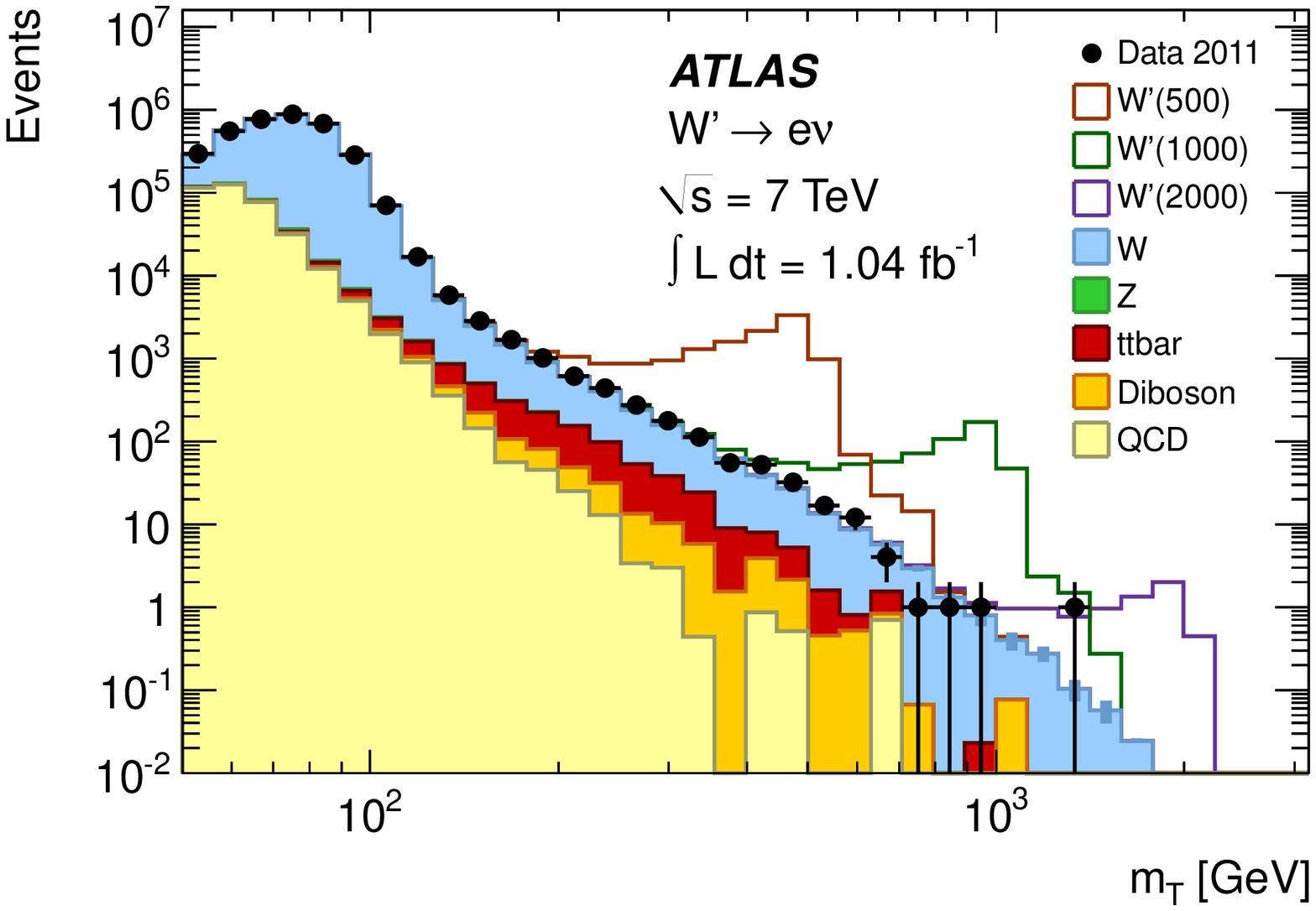}
  \includegraphics[width=0.48\textwidth]{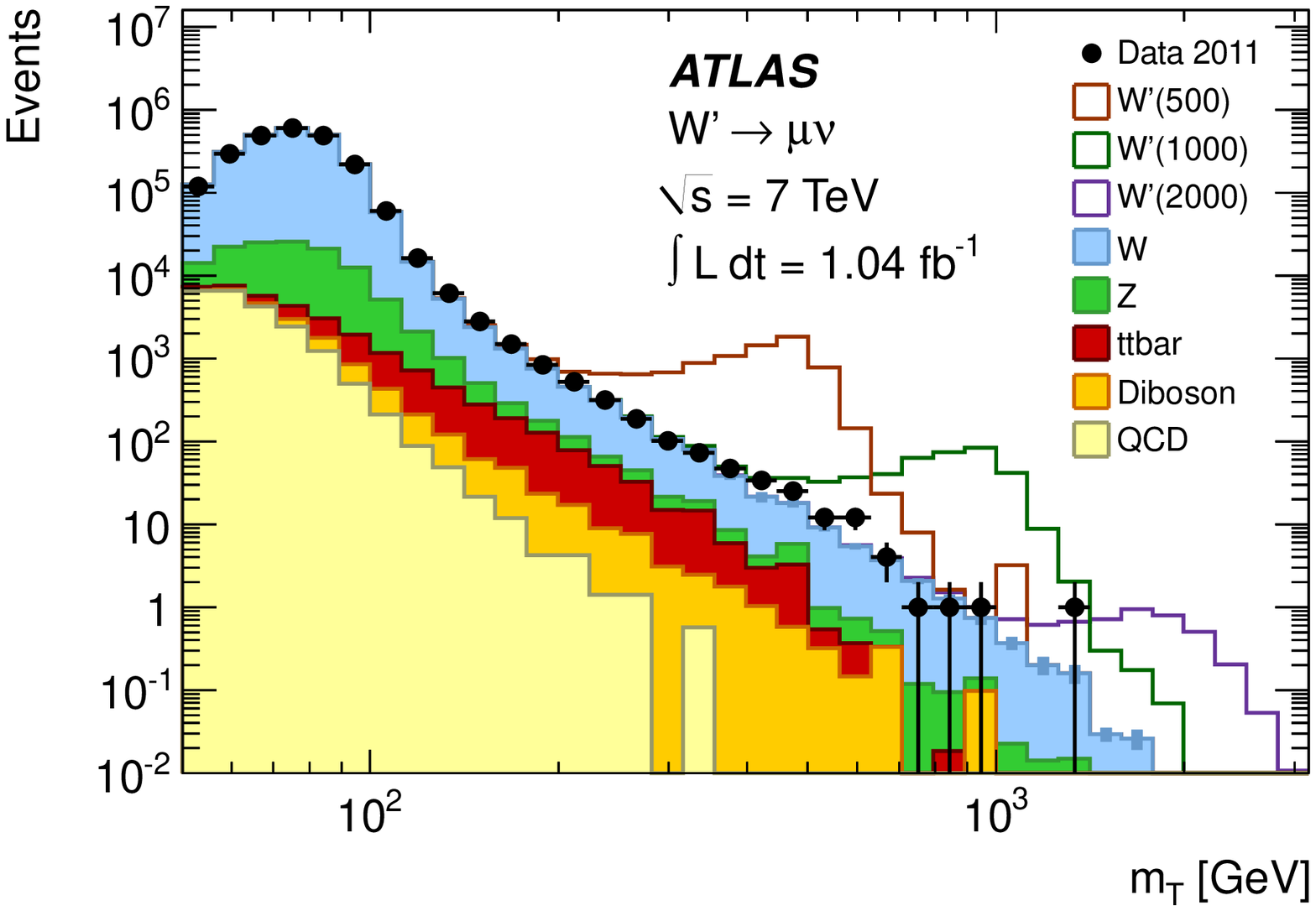}
  \caption{Spectra of \pt\ (top), \mettext\ (center) and \mt\ (bottom)
  for the electron (left) and muon (right) channels after event selection.
  The points represent data and the filled histograms show the stacked backgrounds.
  Open histograms are \wpl\ signals added to the background with masses in \gev\ indicated in parentheses in the legend. 
  The QCD backgrounds estimated from data are also shown.
  The signal and other background samples are normalized using the integrated luminosity of the data
  and the NNLO (approximate-NNLO for \ttbar) cross sections listed in Table~\ref{tab:xsec}.
  \label{fig:final_mt}
  }
\end{figure*}

\begin{table}[!bt]
\caption{
Expected numbers of events in 1.04~\ifb\ from the various background
sources in each decay channel
for \mbox{$\mt>891 \gev$}, the region used to
search for a \wp\ with a mass of 1500~\gev.
The \wlnu\ and \zll\ entries include the expected contributions from the $\tau$-lepton.
No muon events are found in the \ttbar\ sample above this \mt\ threshold.
The uncertainties are statistical.
}
\label{tab:nbg}
\begin{center}
\begin{tabular}{l | l|l}
\hline
\hline
                & \multicolumn{1}{c|}{$e\nu$} & \multicolumn{1}{c}{$\mu\nu$} \\
\hline
\tspace
\wlnu           & 1.59\phantom{000} $\pm$ 0.13         & 1.36\phantom{0}  $\pm$ 0.13 \\
\zll            & 0.00010\phantom{} $\pm$ 0.00004      & 0.095\phantom{}  $\pm$ 0.005 \\
diboson         & 0.08\phantom{000} $\pm$ 0.08         & 0.11\phantom{0}  $\pm$ 0.08 \\
\ttbar          & 0.08\phantom{000} $\pm$ 0.08         & 0 \\
QCD             & 0\phantom{.00000} $^{+0.17}_{-0}$    & 0.01\phantom{0} $^{+0.02}_{-0.01}$ \\
\hline
\tspace
Total           & 1.75\phantom{000} $^{+0.24}_{-0.18}$ & 1.57\phantom{0} $\pm$ 0.15 \\
\hline
\hline
\end{tabular}
\end{center}
\end{table}

\section{Statistical analysis}

A Bayesian analysis is performed to determine if there is significant evidence for existence
of a \wpl\ signal above the SM background and to set limits on that process.
For each candidate mass and decay channel, events are counted above an \mt\ threshold,
$\mt > \mtmin$, with the threshold chosen to maximize sensitivity.
The expected number of events in each channel is
\begin{equation}
\nexp = \effsig \lint \xbr + \nbg,
\end{equation}
where \lint\ is the integrated luminosity of the data sample and \effsig\ is the event selection efficiency,
i.e.\ the fraction of events that pass event selection criteria and have \mt\ above threshold.
\nbg\ is the expected number of background events.
Using Poisson statistics, the likelihood to observe \nobs\ events is
\begin{equation}
  {\cal L}(\nobs|\xbr) = \frac{(\lint \effsig \xbr + \nbg)^{\nobs} e^{-(\lint \effsig \xbr + \nbg)}}{\nobs!}.
\label{eqn:lhood}
\end{equation}
Uncertainties are handled by introducing Gaussian nuisance parameters \npi, each with
a probability density function (pdf) \pdfi, and integrating the product of
the Poisson likelihood with the pdfs.
The integrated likelihood is
\small
\begin{equation}
  {\cal L}_B(\nobs|\xbr) =  \int  {\cal L}(\nobs|\xbr)   \prod \pdfi d\npi.
\label{eqn:lhoodn}
\end{equation}
\normalsize
The nuisance parameters are taken to be the explicit dependencies: \lint, \effsig\ and \nbg,
with the latter evaluated at the central value of \lint.
Correlations between the nuisance parameters are neglected. This is justified by the small effect that the
nuisance parameters themselves have on the limits, as demonstrated below.

The measurements in the two decay channels are combined assuming the same branching fraction for each.
Equation~(\ref{eqn:lhoodn}) remains valid with the Poisson likelihood replaced by the product
of the Poisson likelihoods for the two channels.
The electron and muon integrated luminosity measurements are fully correlated.
The selection efficiencies are uncorrelated and the background levels are partly correlated,
including only the full correlation between the cross section uncertainties in the two channels.
The effect of this correlation is small: if it is not included, the observed \xbr\ limits for
the lowest mass points improve by 2\% and those for the high-mass points are unchanged.

Bayes theorem gives the posterior probability that the \wpl\ has signal strength \xbr:
\begin{equation}
\ppost(\xbr|\nobs) = N {\cal L}_{B}(\nobs|\xbr) \, \pprior(\xbr)
\label{eqn:ppost}
\end{equation}
where $\pprior(\xbr)$ is the assumed prior probability, here chosen to be one (i.e.\ flat in \xbr)
for $\xbr > 0$.
The constant factor $N$ normalizes the total probability to one.
The posterior probability is evaluated for each mass and each decay channel and their combination,
and then used to assess discovery significance and set a limit on \xbr.

\section{Parameter estimation and systematics}

The inputs for the evaluation of ${\cal L}_B$ (and hence \ppost) are \lint, \effsig, \nbg, \nobs\ and the uncertainties on
the first three.
Except for \lint\ and its uncertainty, these inputs are all listed in
Table~\ref{tab:limit_input}.
The uncertainties on \effsig\ and \nbg\ account for simulation statistics and
all relevant experimental and theoretical effects except
for the uncertainty on the integrated luminosity.
The latter is included separately to allow for the correlation between signal and background.
The table also lists the predicted numbers of signal events, \nsig, with their uncertainties accounting
for the uncertainties in both \effsig\ and the cross-section calculation.

\begin{table*}[htbp]
\caption{Inputs for the \wpe\ and \wpmu\ \xbr\ limit calculations.
The first three columns are the \wp\ mass, \mt\ threshold and decay channel.
The next two are the signal selection efficiency, \effsig, and the prediction
for the number of signal events, \nsig, obtained with this efficiency.
The last two columns are the expected number of background
events, \nbg, and the number of events observed in data, \nobs.
The uncertainties on \nsig\ and \nbg\ include contributions from the uncertainties on the
cross sections but not from that on the integrated luminosity.
\label{tab:limit_input}
}
\begin{center}
\begin{tabular}{rrr| r@{ $\pm$ }r r@{ $\pm$ }r r@{ $\pm$ }r r}
\hline
\hline
 \mwp   & \mtmin & \\
~[\gev] & [\gev] &     & \multicolumn{2}{c}{\effsig} & \multicolumn{2}{c}{\nsig} & \multicolumn{2}{c}{\nbg} & \nobs \\
\hline
\hline
\multirow{2}{*}{500} & \multirow{2}{*}{398}
   & $e\nu$   & 0.388 & 0.019 & 6930\phantom{.00} & 620\phantom{.00}  & 101.9\phantom{00} & 10.8\phantom{00}  & 121 \\
 & & $\mu\nu$ & 0.252 & 0.015 & 4500\phantom{.00} & 430\phantom{.00}  &  63.7\phantom{00} &  6.5\phantom{00}  &  91 \\
\hline
\multirow{2}{*}{600} & \multirow{2}{*}{447}
   & $e\nu$   & 0.456 & 0.022 & 3910\phantom{.00} & 330\phantom{.00}  &  62.1\phantom{00} &  7.1\phantom{00}  &  69 \\
 & & $\mu\nu$ & 0.286 & 0.016 & 2450\phantom{.00} & 220\phantom{.00}  &  41.8\phantom{00} &  4.7\phantom{00}  &  57 \\
\hline
\multirow{2}{*}{750} & \multirow{2}{*}{562}
   &   $e\nu$ & 0.429 & 0.020 & 1420\phantom{.00} & 110\phantom{.00}  &  20.7\phantom{00} &  3.7\phantom{00}  &  20 \\
 & & $\mu\nu$ & 0.293 & 0.017 &  970\phantom{.00} &  79\phantom{.00}  &  14.3\phantom{00} &  1.4\phantom{00}  &  20 \\
\hline
\multirow{2}{*}{1000} & \multirow{2}{*}{708}
   &   $e\nu$ & 0.482 & 0.022 &  417\phantom{.00} &  35\phantom{.00}  &   6.13\phantom{0} &  0.92\phantom{0}  &   4 \\
 & & $\mu\nu$ & 0.326 & 0.019 &  282\phantom{.00} &  26\phantom{.00}  &   4.98\phantom{0} &  0.54\phantom{0}  &   4 \\
\hline
\multirow{2}{*}{1250} & \multirow{2}{*}{794}
   &   $e\nu$ & 0.527 & 0.024 &  143\phantom{.00} &  14\phantom{.00}  &   3.09\phantom{0} &  0.49\phantom{0}  &   3 \\
 & & $\mu\nu$ & 0.367 & 0.021 &   99\phantom{.00} &  10\phantom{.00}  &   2.87\phantom{0} &  0.34\phantom{0}  &   3 \\
\hline
\multirow{2}{*}{1500} & \multirow{2}{*}{891}
   &   $e\nu$ & 0.541 & 0.026 &   49.6\phantom{0} &   6.0\phantom{0}  &   1.75\phantom{0} &  0.32\phantom{0}  &   2 \\
 & & $\mu\nu$ & 0.374 & 0.024 &   34.4\phantom{0} &   4.4\phantom{0}  &   1.57\phantom{0} &  0.23\phantom{0}  &   2 \\
\hline
\multirow{2}{*}{1750} & \multirow{2}{*}{1000}
   &   $e\nu$ & 0.515 & 0.024 &   17.3\phantom{0} &   2.4\phantom{0}  &   0.89\phantom{0} &  0.20\phantom{0}  &   1 \\
 & & $\mu\nu$ & 0.338 & 0.020 &   11.4\phantom{0} &   1.7\phantom{0}  &   0.82\phantom{0} &  0.14\phantom{0}  &   1 \\
\hline
\multirow{2}{*}{2000} & \multirow{2}{*}{1122}
   &   $e\nu$ & 0.472 & 0.023 &    6.16\phantom{} &   0.99\phantom{}  &   0.48\phantom{0} &  0.10\phantom{0}  &   1 \\
 & & $\mu\nu$ & 0.323 & 0.021 &    4.21\phantom{} &   0.70\phantom{}  &   0.44\phantom{0} &  0.09\phantom{0}  &   1 \\
\hline
\multirow{2}{*}{2250} & \multirow{2}{*}{1122}
   &   $e\nu$ & 0.415 & 0.019 &    2.84\phantom{} &   0.50\phantom{}  &   0.48\phantom{0} &  0.10\phantom{0}  &   1 \\
 & & $\mu\nu$ & 0.288 & 0.018 &    1.97\phantom{} &   0.36\phantom{}  &   0.44\phantom{0} &  0.09\phantom{0}  &   1 \\
\hline
\multirow{2}{*}{2500} & \multirow{2}{*}{1122}
   &   $e\nu$ & 0.333 & 0.018 &    0.81\phantom{} &   0.16\phantom{}  &   0.48\phantom{0} &  0.10\phantom{0}  &   1 \\
 & & $\mu\nu$ & 0.221 & 0.017 &    0.53\phantom{} &   0.11\phantom{}  &   0.44\phantom{0} &  0.09\phantom{0}  &   1 \\
\hline
\hline
\end{tabular}
\end{center}
\end{table*}

The maximum value for the signal selection efficiency is at $\mwp = 1500\gev$.
For lower masses, the efficiency falls because the relative \mt\ threshold, $\mtmin/\mwp$,
increases to reduce the background level.
For higher masses, the efficiency falls because a large fraction of the cross section goes to
off-shell production with $\mlnu \muchless \mwp$.

The fraction of fully simulated signal events that pass the event selection and are above the \mt\ threshold provides the initial
estimate of \effsig\ for each mass.
Small corrections are made to account for the difference in acceptance at NNLO
(obtained from FEWZ) and that in the LO simulation.
These vary from a 7\% increase for $\mwp = 500\gev$ to a 10\% decrease for $\mwp = 2500\gev$.
Contributions from \mbox{\wptau} with the \taulep\ decaying leptonically have been neglected and would increase the \wp\
event selection efficiencies by 3-4\% for the highest masses.
The background level is estimated for each mass by summing the EW and \ttbar\ event counts
from simulation, and adding the small QCD contribution in the electron channel.

The experimental systematic uncertainties include efficiencies for the electron or muon trigger, reconstruction
and selection.
Lepton momentum and \mettext\ response, characterized by scale and resolution, are also included.
Most of these performance metrics are measured at relatively low \pt\ and their values are extrapolated to
the high-\pt\ regime relevant to this analysis.
The uncertainties in these extrapolations are included but are too small to significantly affect the results.
The uncertainty on the QCD background estimate also contributes to the background level uncertainties for the electron channel.
In some cases, e.g. the \mettext\ scale and the muon QCD background, the experimental systematic uncertainties are
significantly reduced from the previous study~\cite{atlas:wprime_2010_pub} because the additional available data
allow more precise determination.
In other cases they are similar or even larger, but have little effect on the final results.

Table~\ref{tab:syst_summary} summarizes the uncertainties on the event selection efficiencies 
and background levels for the \mbox{\wpl} signal with $\mwp=1500\gev$ using $\mt>891 \gev$.

\begin{table*}[]
\caption{
  Relative uncertainties on the event selection efficiency and background level for
  a \wp\ with a mass of 1500~\gev.
  The efficiency uncertainties include contributions from trigger, reconstruction and event selection.
  The cross section uncertainty for \effsig\ is that assigned to the acceptance correction described in the text.
  The last row gives the total uncertainties.
\label{tab:syst_summary}
}
\begin{center}
\begin{tabular}{l|rr|rr}
\hline
\hline
 & \multicolumn{2}{c|}{\effsig} & \multicolumn{2}{c}{\nbg} \\
 Source                      &  \multicolumn{1}{c}{$e\nu$}  & \multicolumn{1}{c|}{$\mu\nu$} &  \multicolumn{1}{c}{$e\nu$}  & \multicolumn{1}{c}{$\mu\nu$} \\
\hline
 Efficiency                  &       3\%  &       4\%  &       3\%  &       4\%  \\
 Energy/momentum resolution  & \multicolumn{1}{c}{-}
                                          &       2\%  &       3\%  &       1\%  \\
 Energy/momentum scale       &       1\%  &       1\%  &       5\%  &       3\%  \\
 QCD background              & \multicolumn{1}{c}{-} & \multicolumn{1}{c|}{-} &
                                                              10\%  &       1\%  \\
 Monte Carlo statistics      &       3\%  &       3\%  &       9\%  &      10\%  \\
 Cross section (shape/level) &       3\%  &       3\%  &      10\%  &      10\%  \\
\hline
 All                         &       5\%  &       6\%  &      18\%  &      15\%  \\
\hline
\hline
\end{tabular}
\end{center}
\end{table*}

\section{Results}

None of the observations for any mass in either channel or their combination has a significance above three-sigma,
so there is no evidence for the observation of \wpl.
Table~\ref{tab:limits_xbr} and Fig.~\ref{fig:limits_xbr} present the
95\% CL (confidence level) observed limits on \xbr\ for both \wpl\ decay channels and their combination.
The figure also shows the expected limits and the theoretical \xbr\ for an SSM \wp.
The intersection between the central theoretical prediction and the observed limits
provides the 95\% CL lower limit on the mass.
Table~\ref{tab:limits_mass} presents the expected and observed \wp\ mass limits for the electron and muon
decay channels and for the combination of the two channels.
The observed combined mass limit is 2.15~\tev.

\begin{figure}[!htbp]
  \centering
  \includegraphics[width=0.49\textwidth]{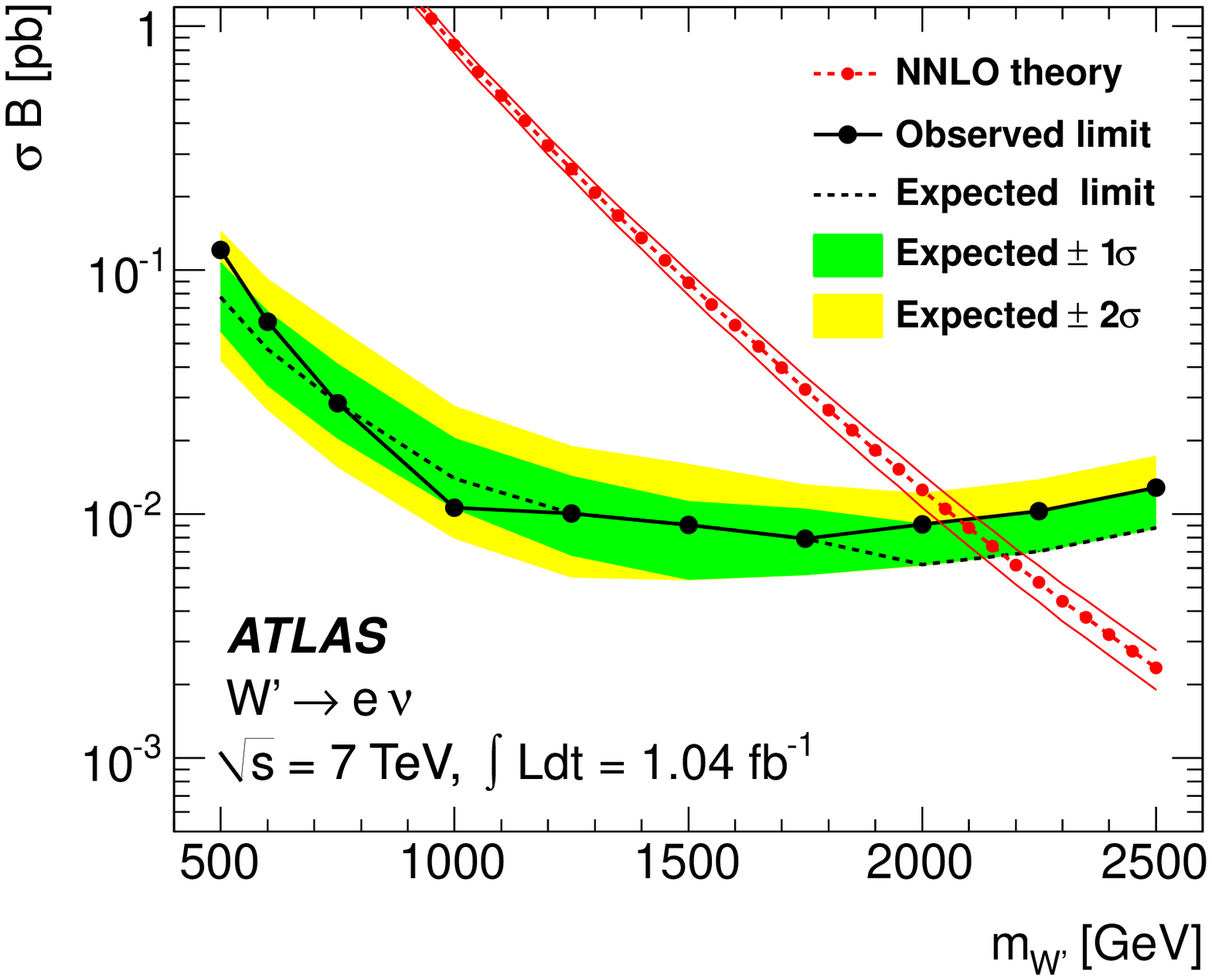}
  \includegraphics[width=0.49\textwidth]{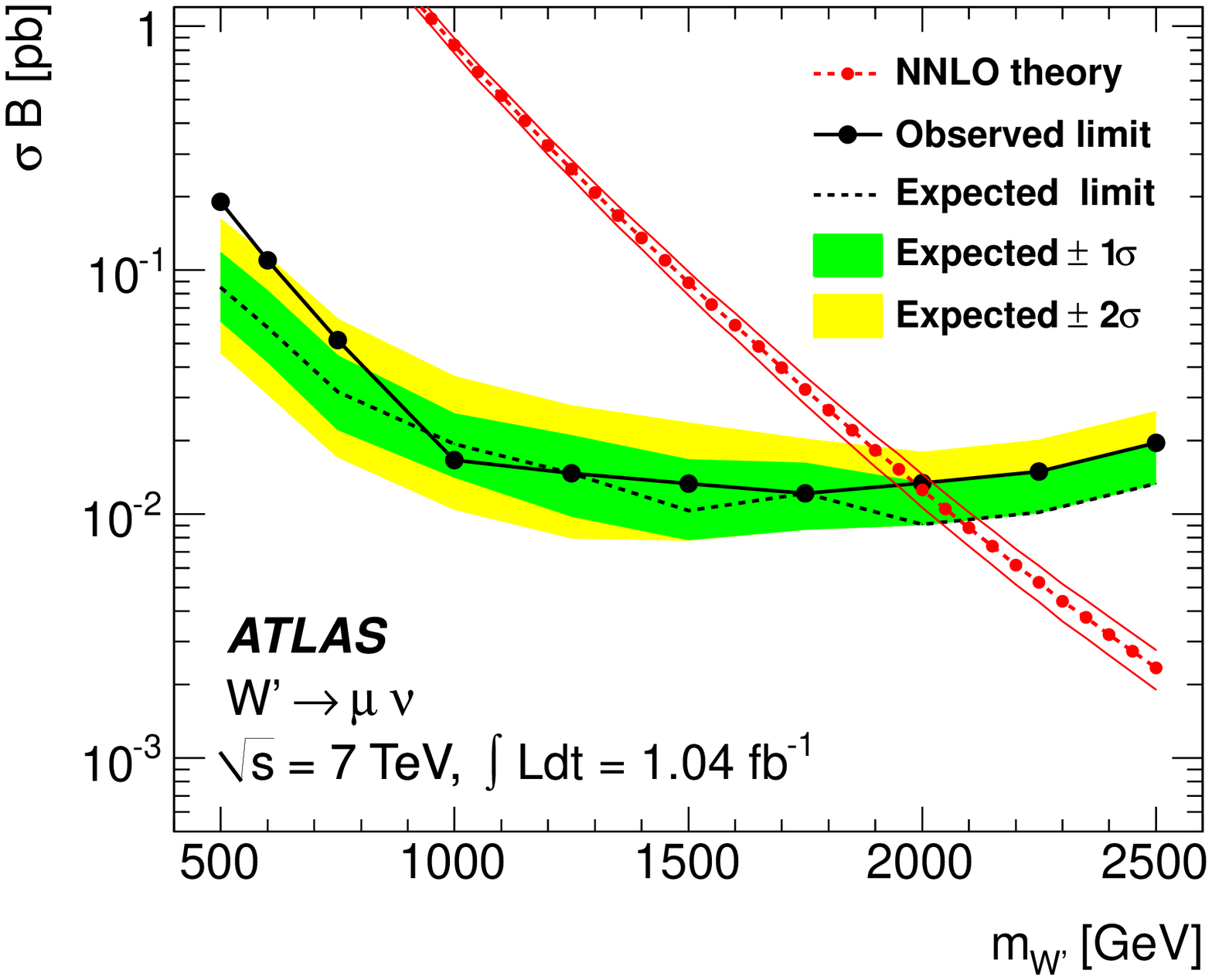}
  \includegraphics[width=0.49\textwidth]{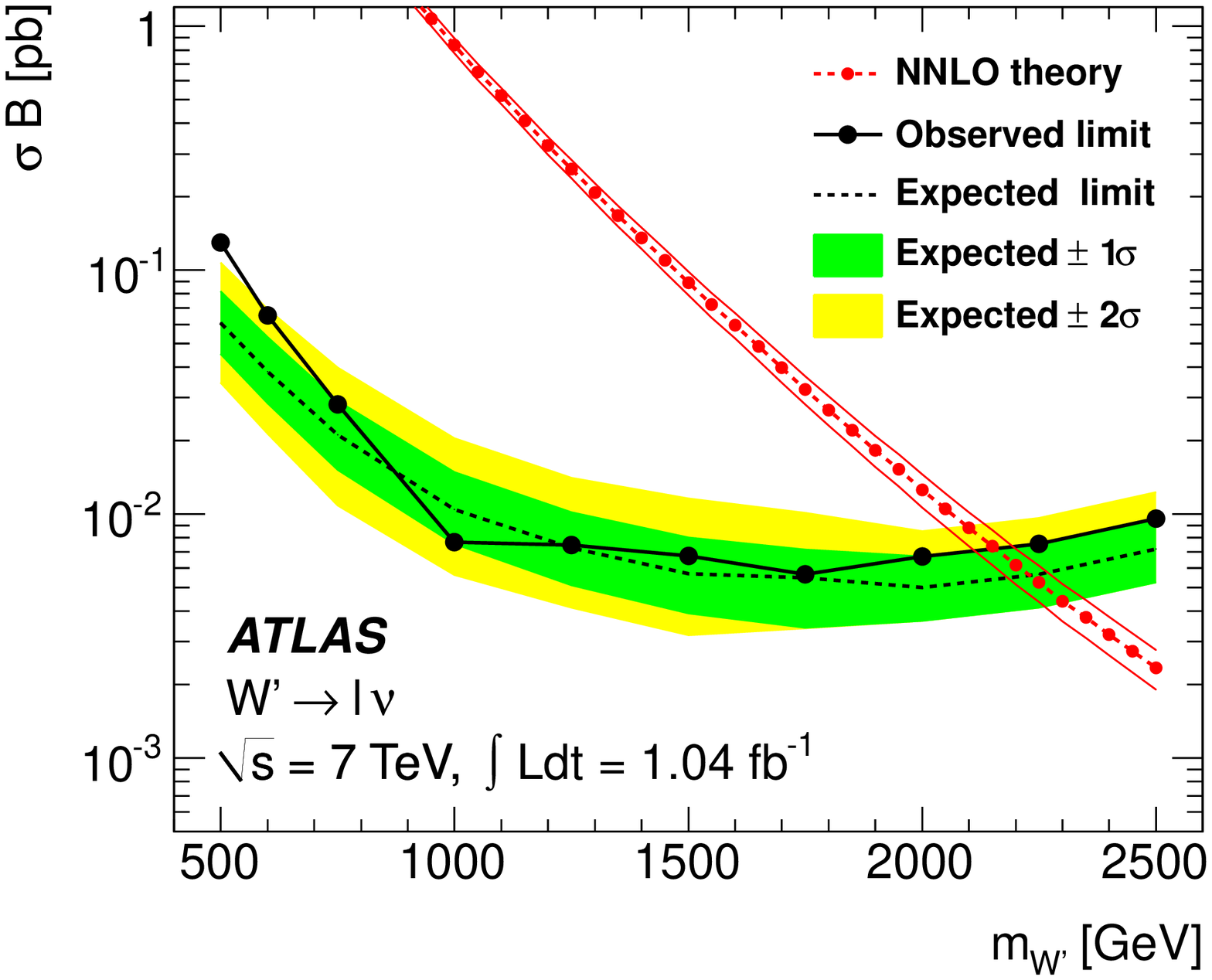}
  \caption{
  Expected and observed limits on \xbr\ for \wpe\ (top), \wpmu\ (center),
  and the combination (bottom) assuming the same branching fraction for both channels.
  The NNLO calculated cross section and its uncertainty are also shown.
  \label{fig:limits_xbr}
  }
\end{figure}

\begin{table}[!htbp]
\caption{Upper limits on \xbr\ for \wpl.
         The first two columns are the mass and decay channel and the following columns are
         the 95\% CL limits with headers indicating the nuisance parameters
         for which uncertainties are included: S for the event selection efficiency (\effsig),
         B for the background level (\nbg), and L for the integrated luminosity (\lint).
         Columns labeled SBL include all uncertainties and are used to evaluate mass limits.
         Results are given for the electron and muon channels and both combined.
\label{tab:limits_xbr}
}
\begin{center}
\begin{tabular}{rc|rrrr}
\hline
\hline
 \mwp    & & \multicolumn{4}{c}{95\% CL limit on \xbr\ [fb]} \\
 ~[\gev] &              & none & S & SB  & SBL \\
\hline
\multirow{3}{*}{500}
 &    $e\nu$ &  97\phantom{.0} &  98\phantom{.0} & 117\phantom{.0} & 121\phantom{.0} \\
 &  $\mu\nu$ & 171\phantom{.0} & 174\phantom{.0} & 186\phantom{.0} & 191\phantom{.0} \\
 &      both & 109\phantom{.0} & 110\phantom{.0} & 127\phantom{.0} & 130\phantom{.0} \\
\hline
\multirow{3}{*}{600}
 &    $e\nu$ &  49\phantom{.0} &  49\phantom{.0} &  59\phantom{.0} &  61\phantom{.0} \\
 &  $\mu\nu$ &  99\phantom{.0} & 100\phantom{.0} & 108\phantom{.0} & 110\phantom{.0} \\
 &      both &  55\phantom{.0} &  55\phantom{.0} &  64\phantom{.0} &  65\phantom{.0} \\
\hline
\multirow{3}{*}{750}
 &    $e\nu$ &  23.0 &  23.1 &  28.1 &  28.5 \\
 &  $\mu\nu$ &  49.2 &  49.8 &  50.9 &  51.7 \\
 &      both &  23.7 &  23.8 &  27.8 &  28.1 \\
\hline
\multirow{3}{*}{1000}
 &    $e\nu$ &  10.1 &  10.2 &  10.5 &  10.6 \\
 &  $\mu\nu$ &  16.1 &  16.3 &  16.5 &  16.7 \\
 &      both &   7.3 &   7.3 &   7.6 &   7.7 \\
\hline
\multirow{3}{*}{1250}
 &    $e\nu$ &   9.8 &   9.9 &  10.0 &  10.1 \\
 &  $\mu\nu$ &  14.4 &  14.5 &  14.6 &  14.7 \\
 &      both &   7.3 &   7.3 &   7.4 &   7.5 \\
\hline
\multirow{3}{*}{1500}
 &    $e\nu$ &   8.8 &   8.9 &   9.0 &   9.0 \\
 &  $\mu\nu$ &  13.0 &  13.2 &  13.2 &  13.3 \\
 &      both &   6.6 &   6.6 &   6.7 &   6.7 \\
\hline
\multirow{3}{*}{1750}
 &    $e\nu$ &   7.8 &   7.9 &   7.9 &   7.9 \\
 &  $\mu\nu$ &  12.0 &  12.1 &  12.1 &  12.2 \\
 &      both &   5.6 &   5.6 &   5.7 &   5.7 \\
\hline
\multirow{3}{*}{2000}
 &    $e\nu$ &   8.9 &   9.0 &   9.0 &   9.1 \\
 &  $\mu\nu$ &  13.2 &  13.3 &  13.3 &  13.4 \\
 &      both &   6.6 &   6.7 &   6.7 &   6.7 \\
\hline
\multirow{3}{*}{2250}
 &    $e\nu$ &  10.2 &  10.2 &  10.3 &  10.3 \\
 &  $\mu\nu$ &  14.8 &  14.9 &  14.9 &  15.0 \\
 &      both &   7.5 &   7.5 &   7.6 &   7.6 \\
\hline
\multirow{3}{*}{2500}
 &    $e\nu$ &  12.7 &  12.8 &  12.8 &  12.9 \\
 &  $\mu\nu$ &  19.2 &  19.5 &  19.6 &  19.7 \\
 &      both &   9.5 &   9.6 &   9.6 &   9.6 \\
\hline
\hline
\end{tabular}
\end{center}
\end{table}

\begin{table}[!htbp]
\caption{Lower limits at 95\% CL on the SSM \wp\ mass.
         The first column is the decay channel ($e\nu$, $\mu\nu$ or both combined) and
         the following columns give the expected (Exp.) and observed (Obs.) mass limits.
\label{tab:limits_mass}
}
\begin{center}
\begin{tabular}{c|rr}
\hline
\hline
        &  \multicolumn{2}{c}{\mwp\ [\tev]} \\
        &  Exp. & Obs. \\
\hline
 $e\nu$    & 2.17 & 2.08 \\
 $\mu\nu$  & 2.08 & 1.98 \\
 both      & 2.23 & 2.15 \\
\hline
\hline
\end{tabular}
\end{center}
\end{table}

The above results are obtained using a prior probability flat in \xbr.
If this prior is replaced by one flat in coupling strength,
the \xbr\ limits improve by 20-28\% for $\mwp \ge 1000\gev$ and
by smaller amounts at the lower masses.
The reference prior~\cite{refprior, hepprior}, which minimizes the information
supplied by the prior, gives intermediate results.
Limits evaluated with \cls~\cite{junk} for the electron and muon channels
and including all uncertainties are nearly identical to the corresponding values
in Table~\ref{tab:limits_xbr}.

Prior to this letter, the best limits for $500 < \mwp < 800\gev$ were established
by CDF~\cite{cdf:Wprime2010} in \wpe\ with \ppbar\ collisions at $\sqrt{s} = 1.96\tev$
using an integrated luminosity of 5.3~\ifb.
At higher masses, the best limits were set by CMS~\cite{cms:wpmu2011mar} and 
ATLAS~\cite{atlas:wprime_2010_pub}, each combining electron and muon channels
and using \pp\ collisions at $\sqrt{s} = 7\tev$ with 36~\ipb\ of data acquired in 2010.
The CDF and CMS limits were obtained with a Bayesian approach,
and the earlier ATLAS results were established with \cls.
Figure~\ref{fig:sigrat_comparison} compares the limits obtained here with those earlier measurements.
The comparison is made using the ratio of the limit to the calculated value of \xbr,
a quantity that is proportional to the square of the coupling strength.
The NNLO cross sections in Table~\ref{tab:xsec} are used for both the ATLAS and CMS points.
The limits presented here provide significant improvement for masses above 600~\gev.

\begin{figure}[!tbp]
  \centering
  \includegraphics[width=0.49\textwidth]{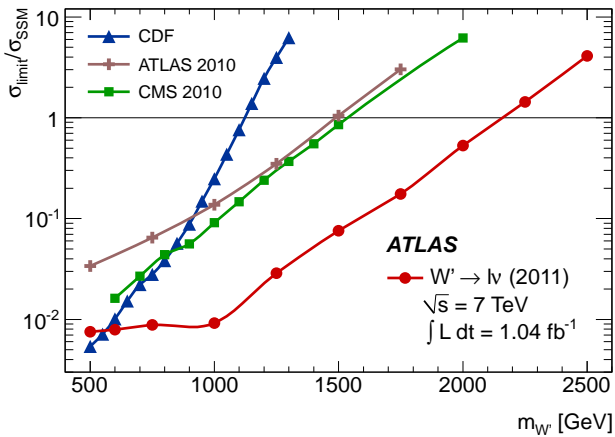}
  \caption{
  Normalized cross section limits ($\sigma_{\rm limit}/\sigma_{\rm SSM}$) for \wpl\
  as a function of mass for this measurement and from CDF~\cite{cdf:Wprime2010},
  CMS~\cite{cms:wpmu2011mar} and the previous ATLAS search~\cite{atlas:wprime_2010_pub}.
  The cross section
  calculations assume the \wp\ has the same couplings as the standard model \w~boson.
  The region above each curve is excluded at the 95\% CL.
  \label{fig:sigrat_comparison}
  }
\end{figure}

\section{Conclusions}

The ATLAS detector has been used to search for new heavy charged gauge bosons decaying to a
lepton plus \mettext.
The search is performed in \pp\ collisions at $\sqrt{s}=7\tev$ using 1.04~\ifb\ of integrated luminosity.
No excess above SM expectations is observed.
Bayesian limits on \xbr\ are shown in Figs.~\ref{fig:limits_xbr} and~\ref{fig:sigrat_comparison}.
These are the best published limits for $\mwp > 600~\gev$.
A \wp\ with SSM couplings is excluded for masses up to 2.15~\tev\ at the 95\% CL.
   
\global\emergencystretch = .2\hsize

\section*{Acknowledgments}

We thank CERN for the very successful operation of the LHC, as well as the
support staff from our institutions without whom ATLAS could not be
operated efficiently.

We acknowledge the support of ANPCyT, Argentina; YerPhI, Armenia; ARC,
Australia; BMWF, Austria; ANAS, Azerbaijan; SSTC, Belarus; CNPq and FAPESP,
Brazil; NSERC, NRC and CFI, Canada; CERN; CONICYT, Chile; CAS, MOST and
NSFC, China; COLCIENCIAS, Colombia; MSMT CR, MPO CR and VSC CR, Czech
Republic; DNRF, DNSRC and Lundbeck Foundation, Denmark; ARTEMIS, European
Union; IN2P3-CNRS, CEA-DSM/IRFU, France; GNAS, Georgia; BMBF, DFG, HGF, MPG
and AvH Foundation, Germany; GSRT, Greece; ISF, MINERVA, GIF, DIP and
Benoziyo Center, Israel; INFN, Italy; MEXT and JSPS, Japan; CNRST, Morocco;
FOM and NWO, Netherlands; RCN, Norway; MNiSW, Poland; GRICES and FCT,
Portugal; MERYS (MECTS), Romania; MES of Russia and ROSATOM, Russian
Federation; JINR; MSTD, Serbia; MSSR, Slovakia; ARRS and MVZT, Slovenia;
DST/NRF, South Africa; MICINN, Spain; SRC and Wallenberg Foundation,
Sweden; SER, SNSF and Cantons of Bern and Geneva, Switzerland; NSC, Taiwan;
TAEK, Turkey; STFC, the Royal Society and Leverhulme Trust, United Kingdom;
DOE and NSF, United States of America.

The crucial computing support from all WLCG partners is acknowledged
gratefully, in particular from CERN and the ATLAS Tier-1 facilities at
TRIUMF (Canada), NDGF (Denmark, Norway, Sweden), CC-IN2P3 (France),
KIT/GridKA (Germany), INFN-CNAF (Italy), NL-T1 (Netherlands), PIC (Spain),
ASGC (Taiwan), RAL (UK) and BNL (USA) and in the Tier-2 facilities
worldwide.

\newpage

%\section*{References}

\bibliographystyle{model1-num-names}
\bibliography{letter}{}

\onecolumn
% Data extracted on 23-Jul-2011 for paperid 112
\begin{flushleft}
{\Large The ATLAS Collaboration}

\bigskip

G.~Aad$^{\rm 48}$,
B.~Abbott$^{\rm 111}$,
J.~Abdallah$^{\rm 11}$,
A.A.~Abdelalim$^{\rm 49}$,
A.~Abdesselam$^{\rm 118}$,
O.~Abdinov$^{\rm 10}$,
B.~Abi$^{\rm 112}$,
M.~Abolins$^{\rm 88}$,
H.~Abramowicz$^{\rm 153}$,
H.~Abreu$^{\rm 115}$
% ...
\bigskip

% ...
$^{10}$ Institute of Physics, Azerbaijan Academy of Sciences, Baku, Azerbaijan\\
$^{11}$ Institut de F\'isica d'Altes Energies and Departament de F\'isica de la Universitat Aut\`onoma  de Barcelona and ICREA, Barcelona, Spain\\
% ...
$^{48}$ Fakult\"{a}t f\"{u}r Mathematik und Physik, Albert-Ludwigs-Universit\"{a}t, Freiburg i.Br., Germany\\
$^{49}$ Section de Physique, Universit\'e de Gen\`eve, Geneva, Switzerland\\
% ...
$^{88}$ Department of Physics and Astronomy, Michigan State University, East Lansing MI, United States of America\\
% ...
$^{111}$ Homer L. Dodge Department of Physics and Astronomy, University of Oklahoma, Norman OK, United States of America\\
$^{112}$ Department of Physics, Oklahoma State University, Stillwater OK, United States of America\\
% ...
$^{115}$ LAL, Univ. Paris-Sud and CNRS/IN2P3, Orsay, France\\
% ...
$^{118}$ Department of Physics, Oxford University, Oxford, United Kingdom\\
% ...
$^{153}$ Raymond and Beverly Sackler School of Physics and Astronomy, Tel Aviv University, Tel Aviv, Israel\\
% ...
\end{flushleft}

\end{document}